# COMPUTATIONAL STUDY OF THE VIBRATING DISTURBANCES TO THE LUNG FUNCTION

**Alexander S. Kholodov**[1], **Sergey S. Simakov**[2],
**Alexey A. Nadolskiy**[3], **Alexander N. Shushlebin**[4]

[1-2] Moscow Institute of Physics and Technology
9, Instituskii Lane, Dolgoprudny, Moscow Region, Russia, 141700
e-mail: [1] xolod@crec.mipt.ru, [2] simakov@crec.mipt.ru,

[3-4] All-Russian Institute of Technical Physics
P.O.Box 245, Snezhinsk, Chelyabinsk Region, Russia, 456770
[3] a.a.nadolsky@vniitf.ru, [4] a.n.shushlebin@vniitf.ru

**Abstract.** *Frequently during its lifetime a human organism is subjected to the acoustical and similar to them vibrating impacts. Under the certain conditions such influence may cause physiological changes in the organs functioning. Thus the study of the oscillatory mechanical impacts to the organism is very important task of the numerical physiology. It allows to investigate the endurance limits of the organism and to develop protective measures in order to extend them. The noise nuisances affects to the most parts of the organism disrupting their functions. The vibrating disturbances caused to the lung function as one of the most sensitive to the acoustical impacts is considered in this work. The model proposed to describe the air motion in trachea-bronchial tree is based on the one dimensional no-linear theory including mass and momentum conservation for the air flow in flexible tubes.*

## 1 INTRODUCTION

Environmental vibrating impacts to the organism of a human play important role for his vital activity. For the most cases such impacts has anthropogenic origin. They affect as in some specific cases (industrial noises generated by different industrial machines and mechanisms) and in everyday life, in particular intensive road traffic in megapolis.

The kind and damage level of the disturbances caused by intensive vibrating impacts result in functioning changes of vestibular analyzer, cardiovascular system, liver and lungs functioning, neurohumoral system and others. Typical symptoms are the headache, coordination of movements disorder, dizziness, symptoms of neurosis-like syndrome and vegetative dysfunction, pain in back muscles and lumbar part of backbone etc. Vibrations may decrease visual acuity, impairment of chromatic sensitivity, loss of functional activity and others. Long-duration acoustical impacts may result in essential changes of blood flow in small vessels of pulmonary circulation thereby reducing oxygen concentration in oxygenated blood. As is well known anoxia give rise to nonreversible processes in vitally important organs. For example risk of pulmonary edema in the presence of anoxia is much higher. Under the intensity of noise impacts exceeding threshold of 100 dB general disorder of organism functioning is observed. In the whole the progress of physiological and pathological changes in organism functioning depends on the nature of the impacts and peculiarities of particular organism.

Thus investigation of vibrating impacts to the human organism is very important task. In particular analysis of the processes causing different physiological and pathological phenomena, forecasting the behavior and looking for the endurance limits of the organism in order to develop protective measures are of great interest. The range of the frequencies up to 20 kHz (acoustical range) is considered in this work as it seems to cause the most crucial effect as it includes eigenfrequencies of most parts of the organism.

One of the parts of the organism most subjected to the acoustical impacts are the lungs. Therefore analysis of the processes excited by the acoustical impacts inside the trachea-bronchial tree and alveolar volume is the final purpose of this work. At the beginning of the paper physiological basis will be discussed that determines mathematical approach to the lungs function modeling. After that the model describing trachea-bronchial tree functioning will be discussed. It will be extended by the model of alveolar volume and the model of aspiration and convective-diffusive substance transfer. Analysis of the acoustical impacts to the nasopharynx and thorax carried out. The results presented in the end of the paper reveal two eigenfrequencies when the thorax is affected and no resonance states for the actions to nasopharynx. Influence of the acoustical impacts to the processes of blood oxygenating will also be discussed. Decrease of oxygen concentration in pulmonary circulation depending on the frequency and amplitude of the disturbances will be evaluated.

## 2 PRELIMINARY DISCUSSION

Respiratory tract similar to the blood vessels constitutes branching structure that maximizes contact area for gas exchange. In accordance with symmetric morphometric model of the lungs inspirited air propagates through the branching tree of bronchial tubes that form dichotomic structure. Trachea-bronchial tree amounts up to 23 generations. Air cells, alveolar ducts and respiratory bronchioles form respiratory and transitional zone amounting 95% of the whole lungs. Conductive zone includes first 16 generations of bronchial tubes where convective component dominates. In other zones diffusion intermixing prevails over the convective transfer [1,2].

Air motion in the most parts of the air tract produce streamline flow. But in some areas in particular near bronchus forks and their pathological narrowing the flow turbulence may oc-cur. Gas composition in alveolar volume supposed to be uniform and gas itself supposed to be incompressible.

The most modern works dealing with lungs mechanics exploit the one-component lung model known from the beginning of XX century. According to this model the lungs are considered as elastic volume having gas-permeable surface. The volume of pleural cavity and air tract is supposed to be negligible with respect to the alveolar volume. Thus thorax volume is supposed to be equal to alveolar one. The air flows into this volume through the system of rigid gas-proof tract usually presented in such models as single channel having some hydraulic resistance. Breathing apparatus in the whole is placed into the elastic cover that is driven by externally defined muscle force. It is supposed that pressure between alveolar volume and thorax depends on time only and is defined by dynamic of respiratory cycle. Such models may explain a lot of experimental data obtained from healthy men. They still used for many problems including complex models using breathing mechanics as one of the elements ([3-15] and others).

An approach proposed in this work extends this model. The main distinction consists in detailed consideration of the conductive zone of the lungs. Trachea-bronchial tree seems to be important part of the breathing mechanics and it taken into consideration using approach developed before for the blood flow modeling in large vessels.

## 3   MODEL OF THE AIR MOTION AND MATTER TRANSPORT

According to specific structure and physiological functioning of different parts of lungs it is proposed to divide mathematical model of air motion into two parts: the model of air motion through conducting zone of the lungs and model of air motion through transitional and respiratory zones.

### 3.1   Air motion in conductive zone of lungs

Basing on the preliminary discussion it seems reasonable to utilize non-stationary model of the flow of viscous incompressible fluid through the hierarchically branching system of elastic tubes. Similar approach has been used in the works [16,17] and others. Such approach is adaptable for our case due to the assumption of flow laminarity.

The basic principles that allow to form the set of equations are the mass and momentum conservation. They may be rewritten for each bronchial tube as:

$$\partial S_k / \partial t + \partial \left( u_k S_k \right) / \partial x = 0 \qquad (1)$$

$$\partial u_k / \partial t + \partial \left( u_k^2/2 + p_k / \rho \right) / \partial x = \psi_k \left( t, x, S_k, u_k, \chi_{ki} \right) \qquad (2)$$

where $t$ — time; $x$ — distance counted from the junction point with the younger bronchial tube; $\rho = 1,23 \cdot 10^{-3} \, g/sm^3$ — gas density; $k$ — index of bronchial tube; $S_k(t,x)$ — cross-sectional area; $u_k(t,x)$ — linear velocity of the flow averaged over cross-section; $p_k(t,x)$ — pressure in the bronchial tube counted off from atmospheric; $\psi_k$ — determine the impact from the external forces (gravitation, friction and others); $\chi_{ki}$ — parameters describing the impact $i$ on bronchial tube $k$.

The domain of calculations for the task (1), (2) is supposed to be elastic cylindrical shell with thin walls. Thus the equation set (1), (2) must be extended with corresponding additional relation determining elastic properties of the shell. This relation may be regarded as "equation of state" for the wall of bronchial tube:

$$p_k - p_{*k} = \rho c_k^2 f_k(S_k) \tag{3}$$

$$f_k(S_k) = \frac{S_k}{S_{0k}} - 1 \tag{4}$$

where $c_k$ — rate of small disturbance propagation; $p_{*k}$ — pressure in the tissues surrounding bronchial tube it also includes the pressure induced by acoustical disturbances affecting the thorax; $S_{0k}$ — mean cross-sectional area of the tube averaged over one respiratory cycle.

At the entrance to the respiratory system the boundary condition was set that is pressure before the trachea:

$$p_1(t,0) = p_T(t), \tag{5}$$

harmonic component must be added to this pressure in the case of acoustical disturbances affecting nasopharynx. Artificial respiration also can be simulated in terms of this model. In this connection constant velocity of air flow at the entrance to the trachea must be specified:

$$u_1(t,0) = u_T(t). \tag{6}$$

Characteristic form of equations (1)-(2) has been used to calculate values at the junction points of respiratory tubes [18-20].

### 3.2 Air motion in transitional and respiratory zones of the lungs

Transitional and respiratory zones of the lungs have a small-scale structure. They consist of highly branching network of small bronchioles and a great number of small alveolar volumes. Hence an approach based on the spatially distributed model is useless in this case. On the other hand transitional and respiratory zones can not be excluded from consideration as they compose up to 95% of the whole volume of the lungs. Thus one-component model of the alveolar volume is proposed to use in this work. Total mass of the lungs and thorax is supposed to be distributed over the surface of the extensible chamber. Mechanical properties of such chamber are defined by integral characteristics including resistance of respiratory tract, air sluggishness, air compressibility, extensibility of the lungs and thorax.

Model used in this work is bases on the equation of ideal gas under isothermal conditions coupled with the motion equation of the chamber walls and linearized integral equation of the air motion through respiratory tract [12]:

$$\left(\sum_{i=1}^{4} a_i dV_{al}/dt\right) + V_{al}/C = p_d(t) - p_g(t) - A\left(Rdp_g/dt + Id^2 p_g/dt^2\right) \tag{7}$$

where $a_1 = (A/C+1)R + R_g$, $a_2 = (A/C+1)I + I_g + ARR_g$, $a_3 = A(RI_g + IR_g)$, $a_4 = AII_g$, $A = 2.46 \cdot 10^{-5}$ [$м^3/кПа$] — ratio of the average volume of the lungs to the average alveolar pressure; $I = 2.5$ [$с^2 \times кПа/м^3$] — air sluggishness; $I_g = 0.21$ [$с^2 \times кПа/м^3$] — sluggishness of the tissues composing respiratory apparatus; $R = 130$ [$с \times кПа/м^3$] — resistance of respiratory tract; $R_g = 110$ [$с \times кПа/м^3$] — resistance of the respiratory tissues; $C = 2.08 \cdot 10^{-4}$ [$м^3/кПа$] — extensibility of the respiratory tissues. At the junction points of pre-alveolar bronchioles with air cells the flow conservation condition must be stated:

$$\sum u_k(t,l_k) \cdot S_k(t,l_k) = \dot{V}_{al}(t) \tag{8}$$

### 3.3 Aspiration and convective-diffusive matter transport

Convective or (and) diffusive matter transport by the flow of air or blood flow has been described using dynamic model [18]:

$$\partial C_{jk} / \partial t + b_{jk} u_k \partial C_{jk} / \partial x = F_{jk}(t, x, C_1, ..., C_J) + \kappa_{jk} \partial^2 C_{jk} / \partial x^2, \quad (9)$$

where $C_{jk}$ — matter concentration designated by $j$ in $k^{th}$ generation of trachea-bronchial or vascular tree; $F_{jk}$ — mass source or leakage, or the rate of matter transformation due to chemical reactions and others; $u_k$ — velocity of the carrier phase (air or blood); $\kappa_{jk}$ — diffusion coefficient; $0 \leq b_{jk} \leq 1$ — coefficient governing convective matter transport (for substances dissolved in blood plasma $b_{jk} = 1$, for macromolecules and regular elements $b_{jk} < 1$).

Boundary conditions for the (9) are similar to the problem of air or blood motion. At the entry to the respiration system boundary conditions are set as value of matter concentration entering during inspiration ($u_1(t,0) > 0$):

$$C_{j1}(t, 0) = C_{jT}(t). \quad (10)$$

For the expiration phase ($u_1(t,0) < 0$) concentration must be obtained from (9). The same equation (9) must be exploited to calculate matter concentration at the branching points of trachea-bronchial tree.

In general case the dependencies of the substance concentrations during propagating through the alveolar septum to the blood are very complex. They defined by different transport processes: membrane transitions, diffusion etc. In simplified case using the assumption of quasi-equilibrium by partial pressures kinetics the values of $C_{j,K1}(t,1)$ and $C_{j,K2}(t,0)$ may be obtained from (9) with the right-hand members:

$$\begin{aligned}
F_{j,K1} &= -\beta_j((p_{ar}(t) + p_0)C_{j,K1}(t,1) - (p_{K2}(t,0) + p_0)C_{j,K2}(t,0)) \\
F_{j,K2} &= \beta_j((p_{ar}(t) + p_0)C_{j,K1}(t,1) - (p_{K2}(t,0) + p_0)C_{j,K2}(t,0)), \beta_j > 0
\end{aligned} \quad (11)$$

where $K1$ and $K2$ — indices of the bronchial tube and corresponded blood vessel. In the case of non-zero diffusion coefficient $\kappa_{jk}$ the condition of the equality of the diffusive flows at the junction points of the trachea-bronchial tree must also be taken into cansideration:

$$\kappa_{jk} \partial C_{jk}(t,1) / \partial x = \kappa_{j,k+1} \partial C_{j,k+1}(t,0) / \partial x. \quad (12)$$

### 4 RESULTS

The models presented in this work have a large number of parameters describing their morphometric and mechanical properties. Some of them are lengths and diameters of respiratory tubes composing trachea-bronchial tree, resistance coefficients, extensibility and sluggishness of the tissues, average volume of lungs and so on. Additional difficulty affecting the models identification arise from the parameters variability depending on sex, age, environmental conditions, physical activity and many others.

The necessary parameters identification was performed using comparison of experimental data and computational results under the normal conditions. In general quite satisfactorily agreement of the computational results with experimental data was achieved. It allows to conclude that the models are adequate enough. After that physiological effects caused by different external disturbances was considered.

### 4.1 Computational study of the resonance properties of the alveolar volume

The simulations carried out in this work dealt with external acoustical disturbances affecting respiratory and blood systems. It supposed that external pressure $p_e$ caused the disturbances of the same frequency $(C_{al} = C_e)$ but different semi-amplitude $(A_{al} = A_e)$ in the alveolar pressure $p_{al}$ (here and later in the paper index $e$ designates parameters of the external impacts while $al$ — designates parameters of alveolar volume). In particular experimental investigations of single-frequency acoustical influences to the dog lungs [25] confirm this assumption. It was shown existence of the eigenfrequency (70 Hz). When affected on this frequency amplitude of the alveolar pressure exceeds the amplitude of affecting pressure by four times. Similar result was obtained from simulations using one-component model of lungs [12] and in the presented work (fig. 1).

Fig. 1 shows that along with the first eigenfrequency of 5 Hz it exist another eigenfrequency equal to 70 Hz for integral alveolar pressure. Mathematical basis for the existence of two eigenfrequencies lays in the fact the differential equation (4) used in the model has the fourth order. Physically this fact may be explained by assigning the major eigenfrequency to the natural vibrations of the lungs tissue while minor eigenfrequency may be assigned to the natural vibrations of the air inside the air-ways.

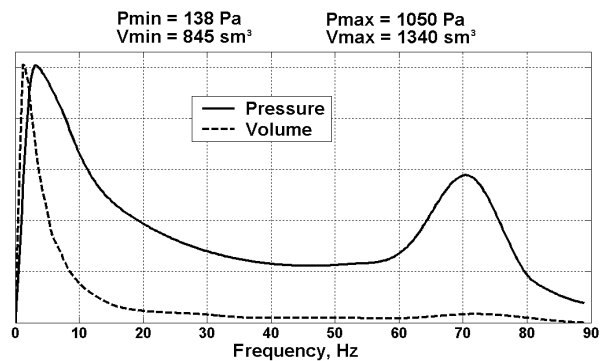

Figure 1: Dependence of the integral values of the alveolar pressure and lungs volume from the frequency of external acoustic disturbances.

### 4.2 Computational study of the resonance properties of trachea-bronchial tree

The following calculations carried out basing on the assumption of symmetrical respiratory cycle that is near to the quiet breathing having ratio of the durations of inspiration and expiration closely to 1.2.

Two types of impacts were analyzed in this work. In the first series of simulations acoustical disturbances affect nasopharynx while in the second thorax was affected. In both cases the semi-amplitude of external impacts was varied in the range from 0 to 0.1 kPa that corresponds to the volume level in the range from 0 to 130-140 dB. Frequency of the external impacts was varied from 0 to 200 Hz. For all simulations total resistance of respiratory system assumed to be constant

$$\sum R_k \approx 2 \; dyne \cdot s / sm^5 .$$

For all figures representing parameters of the respiratory tree the notation of the generations was taken that is collected in table 1.

| Number of generation | Name of the element of trachea-bronchial tree |
|---|---|
| 1 | Trachea |
| 2 | Main lobar bronchus |
| 3 | Main segmental bronchus |
| 4 | Lobular bronchi |
| 5 | Respiratory bronchioles (acinus) |

Table 1: Generations of the respiratory tree notation.

Nomenclature for this figures represented in table 2. All graphs placed one over another on the same picture have the same time scale (in seconds) shown in most cases under the first line of the graphs.

| Notation | Description | Dimension |
|---|---|---|
| $A_t$ | semi-amplitude of disturbances, acting on nasopharynx | $kPa$ |
| $A_e$ | semi-amplitude of disturbances of pleural pressure | $kPa$ |
| $C_e$ | frequency of external acoustic disturbances | $Hz$ |
| k | generation number of tree of respiratory tubes or blood vessels | |
| $N_k$ | amount of respiratory tubes or blood vessels in corresponding generation | |
| $R_k$ | resistance of k-th generation of tree of respiratory tubes or blood vessels | $dyne \cdot s / sm^5$ |
| $C_k$ | propagation velocity of small disturbances | $sm/s$ |
| $L_k$ | length of respiratory tubes or blood vessels of k-th generation | $sm$ |
| $D_k$ | diameter of respiratory tubes or blood vessels of k-th generation | $sm$ |

Table 2: Nomenclature.

As may be observed from fig. 2 amplitude of the pressure oscillations in the bronchial tubes increased under the normal conditions together with generation number increase. The maximum is achieved at the terminal branches connected with alveolar volume. At the same time linear velocity of the inspired air drops from maximum value at the entry of the respiratory system (nasopharnyx) to almost zero value at the junction with alveolar volume. Such behavior is explained by sharp increase of the total cross-sectional area of the trachea-bronchial tree. Volumetric flow remains constant in every generation of the vessels.

In the presence of acoustical disturbances having small frequency (up to 1 Hz) no resonance effects observed as there is no any noticeable discrepancy between the amplitudes of the alveolar and affecting pressure. Such behavior remained the same in the set of numerical experiments for a wide range of parameter $R_k$ and $c_k$.

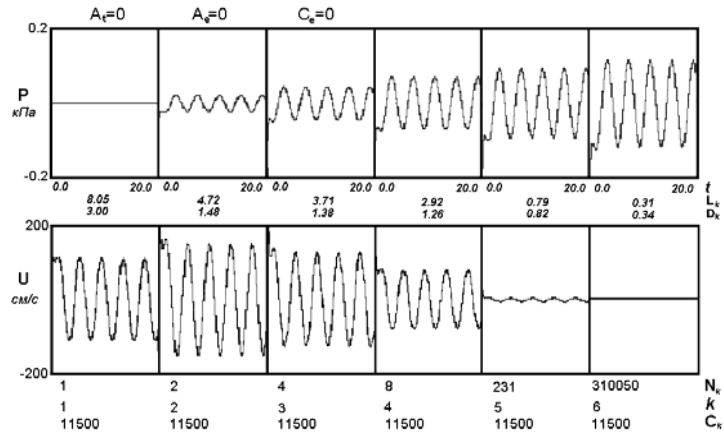

Figure 2: Pressure and linear velocity of the air in trachea-bronchial tree under the normal conditions.

In the range of affecting frequencies from 1 to 10 Hz the maximum of alveolar pressure is observed at the values from 3 to 8 Hz (fig. 3, 4). Specific value of the resonance frequency is defined by the set of parameters $I, I_g, R, R_g, C$ while dependence on the $R_k$ and $c_k$ is rather weaker here. In fact the resonance of the alveolar pressure in the case of pleural pressure disturbances is observed ($A_e \neq 0$, fig. 3, 4) while for the nasopharnyx disturbances no resonance effects were found ($A_e = 0, A_t \neq 0$, fig. 5).

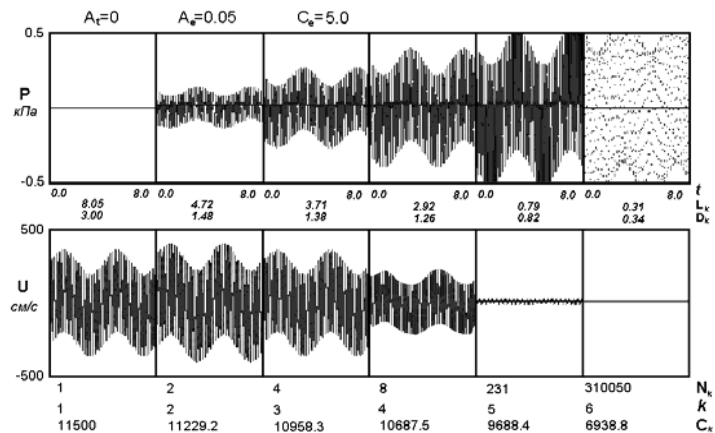

Figure 3: The first maximum of the alveolar pressure amplitude (5 Hz).

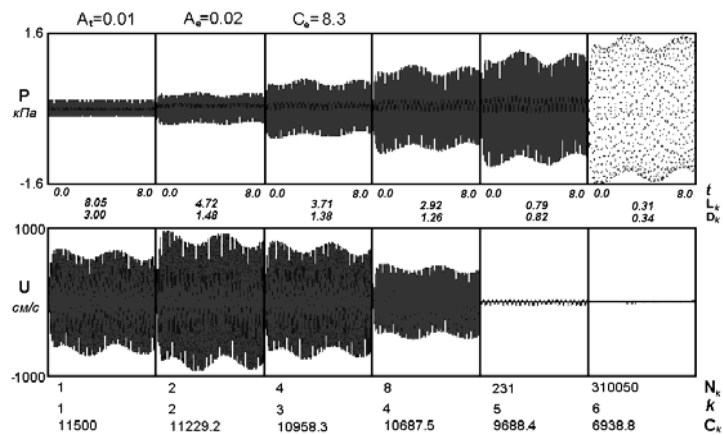

Figure 4: The first maximum of the alveolar pressure amplitude (8.3 Hz).

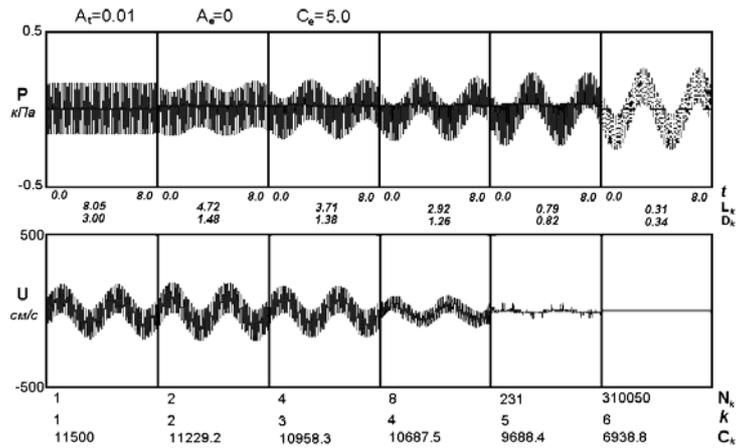

Figure 5: Nasopharnyx acoustical disturbances.

During the further increase of the affecting frequency from 10 to 200 Hz the second resonance maximum of the alveolar pressure amplitude is observed (fig. 6, 7). Specific value of the corresponding frequency as well as resonance response in amplitude of the pressure is greatly depends on the parameters $R_k$ and $c_k$. Under the certain physiologically approved values of these parameters both of eigenfrequencies may coincide.

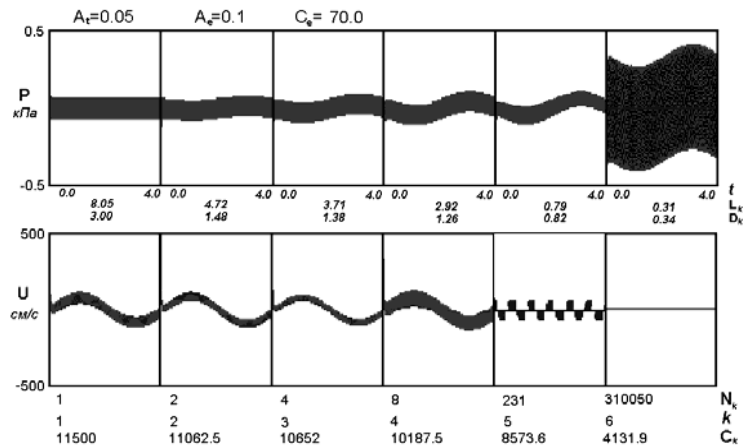

Figure 6: The second maximum of the alveolar pressure amplitude.

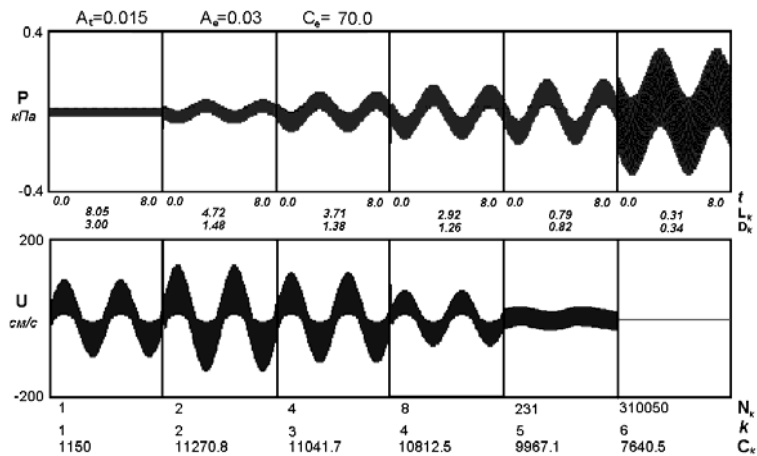

Figure 7: The second maximum of the alveolar pressure amplitude.

Similar to the previous set of simulations the resonance is observed only for the cases when the pleural pressure was affected. For the pressure disturbances at the entry to the trachea-bronchial tree no resonance effects were found (fig. 8).

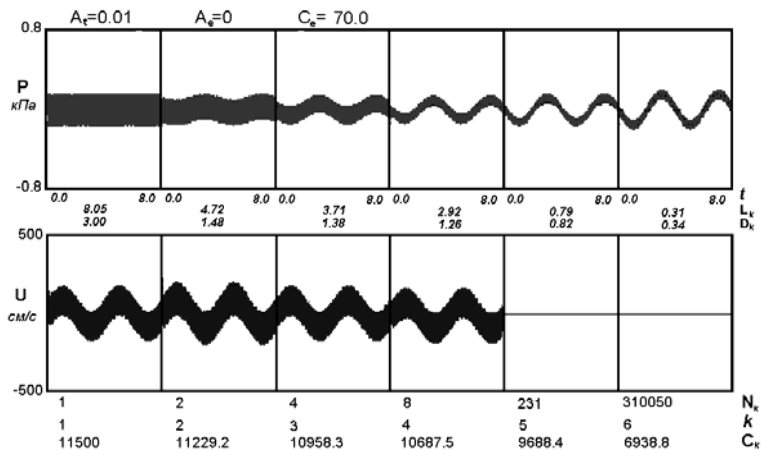

Figure 8: Influence of acoustic disturbances on nasopharynx.

During the further increase of the affecting frequency excitation of oscillations in large bronchi was observed while amplitude of the pleural pressure is greatly decreased (fig. 9, 10).

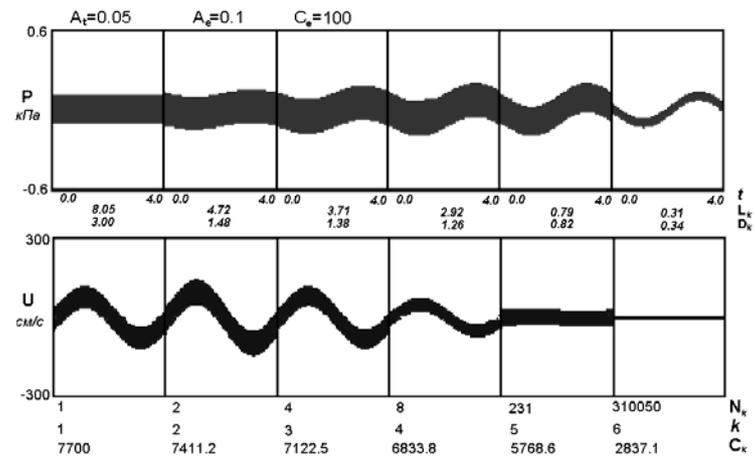

Figure 9: Parameters of trachea-bronchial tree under non-resonance frequencies of impact.

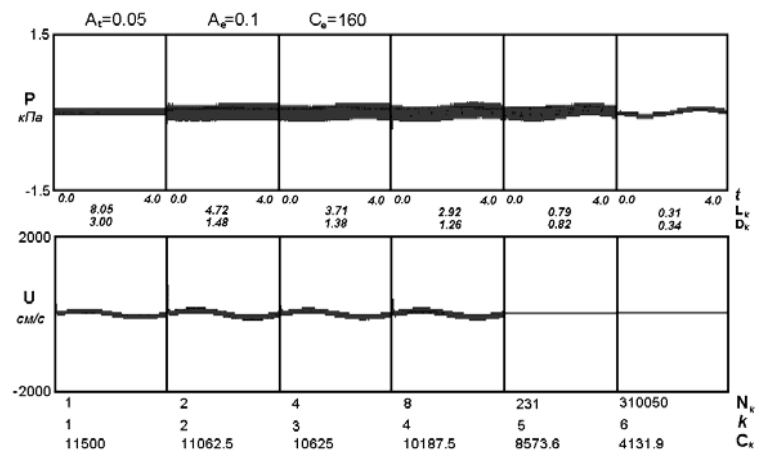

Figure 10: Parameters of trachea-bronchial tree under non-resonance frequencies of impact.

### 4.3 Computational study of the resonance properties of pulmonary circulation

It turns out that acoustical disturbances of the lungs discussed above significantly affect hemodynamics of the pulmonary circulation. The earlier developed dynamical model of the flow of viscous incompressible fluid [18-20] allowed us to simulate this effect. To do this the models of the air and blood flows were joined together by means of common boundary conditions. They were set in accordance with assumption that interaction between respiratory and cardiovascular system is realized through the alveolar pressure acting to the walls of the small vessels only. Since the alveolar pressure is mostly affected by the impacts having resonance frequencies (fig. 3, 4, 6, 7) the influence of the same impacts is of particular interest here.

The results of simulations reveal that when affected by impacts exciting resonance in alveolar pressure the hemodynamics in the small vessels of pulmonary circulation undergo the most changes. In particular blood flow in these vessels becomes alternating (fig. 11). Such result is somewhat obvious if we recall that these vessels are directly affected by the alveolar pressure. At the same time in the vessels of the 3-5 generations above and especially below micro-vessels (venous part) substantial changes in blood flow are also observed. For the other affecting frequencies (from 0 to 200 Hz) the disturbances of the blood flow were not observed.

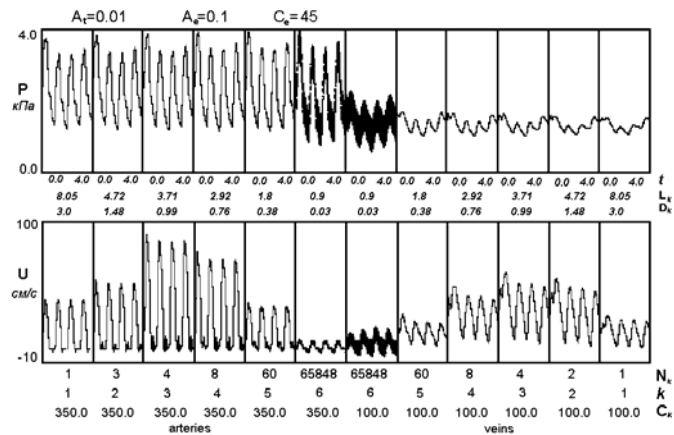

Figure 11: Resonance of the pressure in pulmonary circulation.

Explanation of these effects lies in the fact that additional pressure gradient is induced by the disturbed alveolar and pleural pressures. This gradient is responsible for the additional pulse waves in the circulatory system that propagating through the vascular network caused substantial changes in the whole cardiovascular system. On a large scale effect is analogous to the heart disorders when pressure profile at the heart exits is disturbed (e.g. arrhythmia).

### 4.4 Computational analysis of the aspiration and convective-diffusive matter transport during acoustical impacts

Models of respiration and blood circulation mentioned above allow to investigate the processes of convective-diffusive matter transport under the normal conditions and in the presence of external acoustical impacts. The results of simulations using (9) are presented in fig. 12-14.

In the first series of simulations the foreign substance penetration into the blood is considered. The substance concentration was set to 10% at the entry to the respiratory system and 0% in the blood at the beginning of the computational experiment. From the fig. 12 it is easy to observe substantial changes in fraction and time profile of the concentration in the blood during the intensive acoustical impacts. The effect is conditioned by substantial changes in blood flow profile mentioned above (fig. 11).

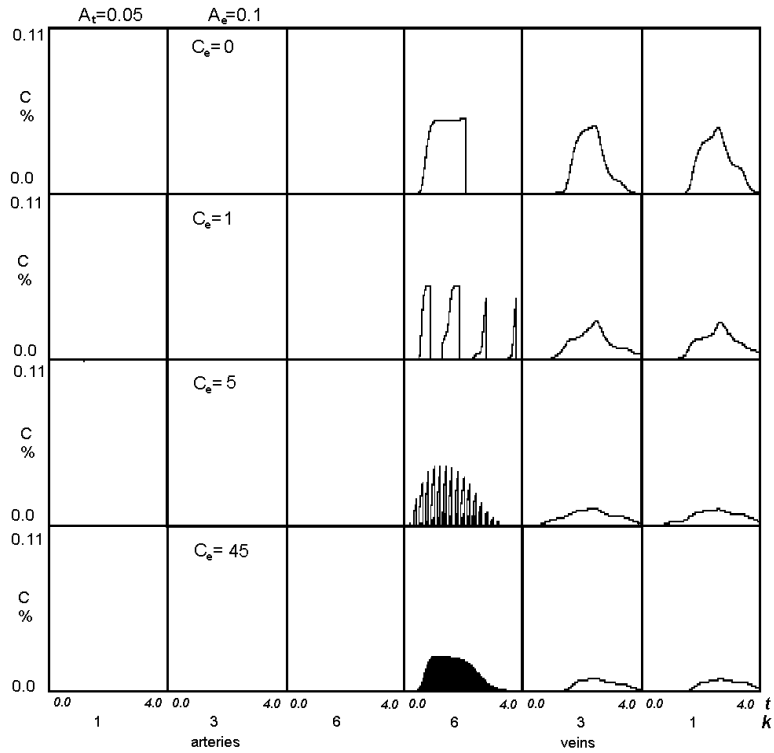

Figure 12: Transport of the foreign substance under the intensive acoustical disturbances.

Similar results for the oxygen and carbon dioxide concentrations in pulmonary blood are depicted in fig. 13. Unlike the previous case initial concentrations was set to 21% and 14% at the entry to the respiratory system.

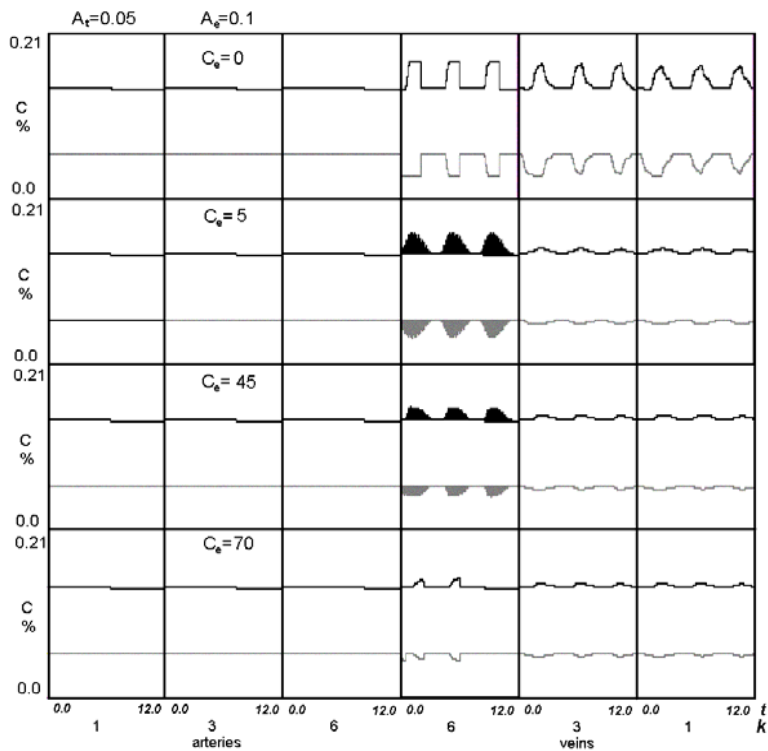

Figure 13: Transport of oxygen (black) and carbon dioxide (grey) under the intensive acoustical disturbances.

The most interesting are the results investigating lungs gas exchange under high intensive (110-130 dB) and rather long-durational (up to 5 minutes) acoustical disturbances. As one may observe from fig. 14 there is essential drop in oxygen concentration during the time typical for the processes of saturation and excretion (approximately 7 minutes for the human).

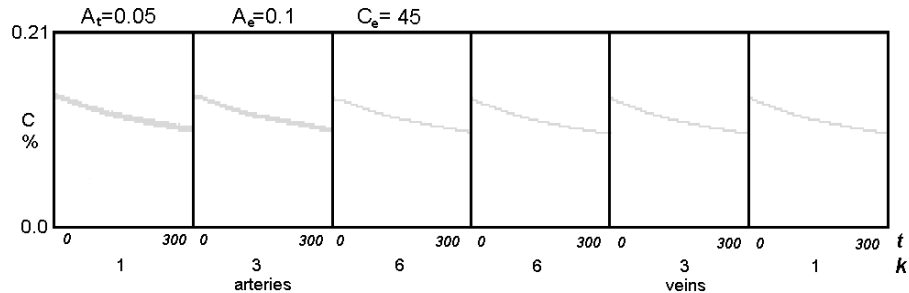

Figure 14: Oxygen concentration in the pulmonary circulation under the high intensive and long-durational acoustical disturbances.

Analysis of the dependence of oxygen concentration in pulmonary circulation form the amplitude of the affecting impacts having frequency 45 Hz is depicted in fig. 15. It is obvious that substantial decrease in concentration takes place much earlier (20-40 Pa) than the pain threshold is achieved (100 Pa).

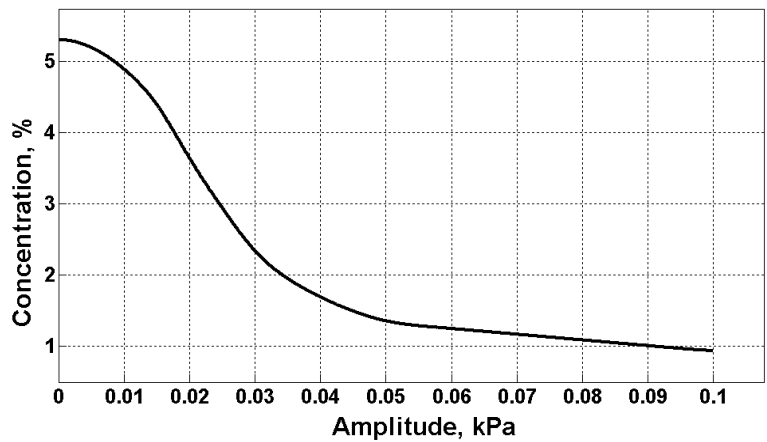

Figure 15: Oxygen concentration in the venous part of pulmonary circulation (45 Hz).

## 5  CONCLUSIONS

- The model of non-stationary flow of viscous incompressible fluid through the set of elastic tubes combined with one-component model of the alveolar volume was adapted to describe respiratory tract mechanics.

- The carried out simulations reveal substantial influence of the acoustical impacts to the air and blood flow dynamic profiles. Noticeable decrease of oxygen concentration in pulmonary circulation was observed much earlier than the pain threshold is achieved. It should be mentioned that specific parameters (affecting amplitudes and eigenfrequencies) depend of course on particular case but qualitative behavior is supposed to be the same.

- The proposed method to combine the models of respiratory and cardiovascular systems is effective and useful for practical applications. In the future it can be extended and adapted for many different tasks of computational biomechanics.

**REFERENCES**

[1] R.F. Schmidt, G. Thews eds., *Human Physiology*. Mir, Moscow, **2**, 1996.

[2] E.B. Babskii, A.A. Zubkov, G.I. Kositskii, B.I. Khodorov, *Human Physiology*. Medicina, Moscow, 1966.

[3] M.Y. Jaeger, A.B. Otis, Effects of compressibility of alveolar gas on dynamics and work of breathing. *J. Appl. Physiol.*, **19**(1), 1964.

[4] I.S. Amosov, V.A. Degtyaryov, V.S. Volkov, D.U. Mendesheva, Biomechanics of thorax under physical load and at rest. *Biomechanics of blood circulation, respiratory and biological tissues*, Riga, 1981.

[5] G.A. Lyubimov, Mechanics of respiratory apparatus. *Biomechanics of blood circulation, respiratory and biological tissues*, Riga, 1981.

[6] V.G. Shabelnikov, Influence of the air flow velocity during inspiration and expiration to the power inputs and pulmonary gas exchange of human organism, *Biomechanics of blood circulation, respiratory and biological tissues*, Riga, 1981.

[7] R. Peslin, C. Duvivier, B. Hannhart, Respiratory mechanical impedances. *Methodology and interpretation,* Biorheology, **1**, 183-191, 1984.

[8] A.C. Jackson, M. Tabrizi, J.W. Watson, M.I. Kotlicoff, Oscillatory mechanics in the dog lung: 4-64 Hz. *Proc. of the 36th ACEMB,* Bethesda, **25**, 209, 1983.

[9] T.K. Miller, R.L. Pimmel, Standard errors on respiratory mechanical parameters, obtained by forced random excitation. *IEEE Trans. Biomed. Eng.*, **30**(12), 826-832, 1983

[10] J.G. Eyles, R.L. Pimmel, Estimating respiratory mechanical parameters in parallel compartment models. *IEEE Trans. Biomed. Eng.* **28(4)**, 313-317, 1981.

[11] H. Felkel, M. Jirina, M. Adamec, et. al., Simulation of human reactions under extreme conditions. *Acta Astronautica*, **8**(9-10), 971-976, 1981.

[12] A.I. Dyachenko, The study of one-component model of lung mechanics. *Medical biomechanics*, Riga, **1**, 147-152, 1986.

[13] G.A. Lyubimov, Study of respiratory mechanics using two-component model including gas compressibility. *II All-Union conference on biomechanical problems*, Riga, 1979.

[14] A.I. Dyachenko, V.G. Shabelnikov, Mathematical model of gravitation distribution of ventilation and blood flow in human lungs. *Biomechanics of blood circulation, respiratory and biological tissues*, Riga, 1981.

[15] Q. Grimal, S. Naili, A. Watzky, A high-frequency lung injury mechanism in blunt thoracic impact. *J. of Biomech.*, Elsevier, **38**, 1247-1254, 2005.

[16] T.J. Pedley, R.C. Schroter, M.F. Sudlow, Energy losses and pressure drop in models of human airways. *J. Respir. Physiol.*, **9**, 371-386, 1970.

[17] J.J. Shin, D. Elad, R.D. Kamm, Simulation of forced breathing maneuvers. *Biological flow*, N.Y. Planum Press, 287-313, 1995.

[18] A.S. Kholodov, Some dynamical models of external breathing and blood circulation taking into account their interaction and matter transfer. *Mathematical models and medicine progress*, Nauka, Moscow, 121-163, 2001.

[19] A.S. Kholodov, S.S. Simakov, A.V. Evdokimov, Y.A. Kholodov, Numerical simulations of cardiovascular diseases and global matter transport. *Proc. of the International Conference Advanced Information and Telemedicine Technologies for Health*, S. Ablameyko et. al. eds., Minsk, **2**, 188–192, 2005.

[20] A.S. Kholodov, S.S. Simakov, A.V. Evdokimov, Y.A. Kholodov, Matter transport simulations using 2D model of peripheral circulation coupled with the model of large vessels. *Proc. of II International Conference On Computational Bioeng.*, H. Rodrigues et. al. eds., IST Press, **1**, 479-490, 2005.

[21] K.M. Magomedov, A.S. Kholodov, *Grid-characteristic numerical methods*. Nauka, Moscow, 1988.

[22] O.V. Vorobyov, Y.A. Kholodov, On the method of numerical integration of 1D problems in gas-dynamics. *J. Math. Mod.*, **8(**1), 77-92, 1996.

[23] A.S. Kholodov, Monotonic difference schemes on irregular grids for elliptic equations in domains with multiple boundaries. *J. Math. Mod.*, **3(**9), 104-113, 1991.

[24] A.S. Kholodov, About majorized differential schemes on irregular grids for hyperbolic equations. *Math. Mod.*, Moscow, MSU, 105-113, 1993.

[25] J.J. Fredberg, Mechanics of the lung during high frequency ventilation. *Proc. of the 36th ACEMB*, Bethesda, **25**, 33, 1983.